\documentclass[aps,prd,showpacs,floatfix,preprintnumbers,groupedaddress]{revtex4}
\usepackage{graphicx}
\usepackage{psfrag}
\def\lsim{\raise0.3ex\hbox{$\;<$\kern-0.75em\raise-1.1ex\hbox{$\sim\;$}}}
\def\gsim{\raise0.3ex\hbox{$\;>$\kern-0.75em\raise-1.1ex\hbox{$\sim\;$}}}

\def\hbar{\hspace{0pt}\raisebox{1pt}{$-$} \hspace{-7pt} h}

\newcommand{\be}{\begin{equation}}
\newcommand{\ee}{\end{equation}}
\newcommand{\bd}{\begin{displaymath}}
\newcommand{\ed}{\end{displaymath}}
\newcommand{\bea}{\begin{eqnarray}}
\newcommand{\eea}{\end{eqnarray}}

\newcommand {\ignore}[1]{}

\def\10{SO(10)}
\def\321{SU(3) $\otimes$ SU(2) $\otimes$ U(1) }

\begin{document}

\title{Searching supersymmetry at the LHCb with displaced vertices}
\date{\today}

\author{F.\ de Campos}
\email{camposc@feg.unesp.br}
\affiliation{Departamento de F\'{\i}sica e Qu\'{\i}mica,
             Universidade Estadual Paulista, Guaratinguet\'a -- SP,  Brazil }

\author{O.\ J.\ P.\ \'Eboli}
\email{eboli@fma.if.usp.br}
\affiliation{Instituto de F\'{\i}sica,
             Universidade de S\~ao Paulo, S\~ao Paulo -- SP, Brazil.}

\author{M.\ B.\ Magro} \email{magro@fma.if.usp.br}
\affiliation{Faculdade de Engenharia,
             Centro Universit\'ario Funda\c{c}\~ao Santo Andr\'e,
             Santo Andr\'e -- SP, Brazil.}

\author{D.\ Restrepo} \email{restrepo@uv.es}
\affiliation{Instituto de F\'{\i}sica, Universidad de Antioquia - Colombia}


\begin{abstract}
\vspace*{1cm}

Supersymmetric theories with bilinear R--parity violation can give
rise to the observed neutrino masses and mixings. One important
feature of such models is that the lightest supersymmetric particle
might have a sufficiently large lifetime to produce detached
vertices. Working in the framework of supergravity models we analyze
the potential of the LHCb experiment to search for supersymmetric
models exhibiting bilinear R--parity violation.  We show that the LHCb
experiment can probe a large fraction of the $m_0 \otimes m_{1/2}$,
being able to explore gluino masses up to 1.3 TeV. The LHCb discover
potential for this kind of models is similar to the ATLAS and CMS ones
in the low luminosity phase of operation of the LHC.

\end{abstract}

\maketitle


\newpage

\section{Introduction}

Recently neutrino physics not only has firmly established a signal of
physics beyond the Standard Model (SM) but also provided us valuable
precise information on neutrino masses and mixings.  Extensions of the
Standard Model should address the neutrino mass generation mechanism,
being able to accommodate the present neutrino knowledge. On the other
hand, weak scale supersymmetry (SUSY) is a very popular solution for
the hierarchy problem which also presents some pleasant features like
the unification of the gauge couplings and the fact of being
perturbative.  R--parity conservation is an {\em ad-hoc} working
hypothesis in SUSY models which can be violated. Moreover, it has been
shown that SUSY models with R--parity non--conservation can generate
neutrino masses and mixings, in particular, models with bilinear
R-parity violation provide an economical effective model for
supersymmetric neutrino
masses~\cite{Diaz:1997xc,chun:1998gp,Diaz:2003as,Hirsch:2000ef,DeCampos:2001wq,deCampos:2005ri}.

It is an experimental fact that supersymmetry is broken in the observed
universe, however, little is known on how this break takes place. In
this work we assume, for the sake of simplicity, gravity mediated
supersymmetry breaking.  Although R--parity violation can be
introduced in a variety of
ways~\cite{hall:1984id,ross:1985yg,ellis:1985gi,marek:1996,masiero:1990uj} we
work in the scenario of bilinear R--parity violation (BRpV). We call this
model BRpV--mSUGRA. One nice feature of this model is that it is
falsifiable at the CERN Large Hadron Collider despite the smallness of
R--parity violation needed to accommodate the neutrino oscillation
data~\cite{Hirsch:2004he,deCampos:2007bn}.

In BRpV--mSUGRA models the lightest supersymmetric particle (LSP) is
no longer stable due to R--parity violating interactions. Moreover,
the LSP can be rather long lived due to the smallness of the BRpV
parameter needed to fit the neutrino masses and
mixings~\cite{deCampos:2007bn,Magro:2003zb}. This opens a new window
for SUSY studies, that is, the search for detached vertices associated
to the decay of rather heavy particles. Although the LHCb experiment
is not able to search for SUSY in the canonical signatures due to its
non--hermicity, it does have good vertexing capabilities which allow
for the SUSY searches in R--parity violating
scenarios~\cite{deCampos:2007bn,Kaplan:2007ap}. Furthermore, the
predicted BRpV--mSUGRA displaced vertices possess a visible invariant
mass much larger than any SM particle, therefore, the search for
displaced vertices is essentially background free.

In this work we study the potential of the LHCb experiment to search
for BRpV--mSUGRA, showing that it can explore a large fraction of the
$m_0 \otimes m_{1/2}$ plane.  At small $m_0$, the LHCb experiment will
be able to study $m_{1/2}$ as high as 600 GeV which corresponds to
neutralino masses of the order of 250 GeV and 1.3 TeV gluino masses.
For $m_{1/2}=500$ GeV the LHCb has a large $m_0$ reach which extends
to $\simeq 2$ TeV. It is interesting to notice that the LHCb reach is
just $\approx 30$\% smaller than the ATLAS and CMS ones in the initial
operation of the LHC, that is, in the low luminosity phase.

Previously, the search for SUSY via displaced vertices at the LHCb has
been studied in Ref.~\cite{Kaplan:2007ap}, however, this work
considered explicit R--parity breaking via baryon number violating
operators ($\lambda^{\prime \prime}$), which leads to pure hadronic
neutralino decays $\tilde{\chi}^0_1 \to q q q$.  Our BRpV-mSGURA model
differs from that scenario in many ways since the R--parity violating
parameters are rather constrained by neutrino physics, which, in turn,
restrict the branching ratios and decay lengths of the LSP. Moreover,
the lightest neutralino does not possess pure hadronically decay
modes, decaying into pure leptonic channels or in semileptonic modes;
see Section II for further details.  Since the LHCb experiment has
already designed a dimuon trigger, it should be straightforward to
look for $\tilde{\chi}^0_1 \to \nu \mu^+ \mu^-$ events originated from
our BRpV--mSUGRA model.

This paper is organized as follows. In Sect.\ \ref{model} we present
the BRpV--mSUGRA model and its main properties. The details of our
analysis, our results and conclusions are in Sect.\ \ref{results}.

\section{Model properties}
\label{model}

The superpotential for the supergravity model with bilinear R-parity
violation is~\cite{Diaz:1997xc}
\begin{equation}
W_{\text{BRpV}} = W_{\text{MSSM}}  + \varepsilon_{ab}
\epsilon_i \widehat L_i^a\widehat H_u^b \; ,
\end{equation}
where $W_{\text{MSSM}}$ is the usual potential of the minimal
supersymmetric standard model and the additional bilinear contribution
contains three parameters ($\epsilon_i$), one for each fermion
generation. The BRpV--mSUGRA model also possesses new soft
supersymmetry breaking terms
\begin{equation}
V_{\text{soft}} = V_{\text{mSUGRA}} - \varepsilon_{ab} 
B_i\epsilon_i\widetilde L_i^aH_u^b
\end{equation}
with three new free parameters ($B_i$).  It is interesting to notice
that there is no field redefinition such that the BRpV terms of the
superpotential and the associated soft terms can be eliminated
simultaneously. Moreover, the bilinear R--parity violating
interactions generate a vacuum expectation value for the sneutrino
fields ($v_i$) after the minimization of the full scalar potential.

In total, the BRpV--mSUGRA model has eleven free parameters which are
\begin{equation}
m_0\,,\, m_{1/2}\,,\, \tan\beta\,,\, {\mathrm{sign}}(\mu)\,,\, 
A_0 \,,\, \epsilon_i \: {\mathrm{, and}}\,\, B_i\,,
\end{equation}
where $m_{1/2}$ and $m_0$ are the common gaugino mass and scalar soft
SUSY breaking masses at the unification scale, $A_0$ is the common
trilinear term, and $\tan\beta$ is the ratio between the Higgs field
vacuum expectation values.

In the BRpV--mSUGRA model neutralinos and neutrinos mix via the BRpV
interactions and through the induced sneutrino vacuum expectation
values, giving rise to a $7 \times 7$ neutral fermion mass matrix.  At
tree level these models exhibit just one massive neutrino at the
atmospheric scale while solar mass scale comes into play via one--loop
radiative corrections
~\cite{Hirsch:2000ef,Diaz:2003as,Hirsch:2004he,Dedes:2006ni}.  It has
been shown in Refs.~\cite{Diaz:2003as,Hirsch:2000ef} that the neutrino
masses and mixings stemming from the diagonalization of the neutral
fermion mass matrix are approximately given by
\begin{equation}
 \begin{array}{ll}
 \Delta m_{12}^2\propto |\vec{\epsilon}| \qquad ;
&\Delta m_{23}^2\propto |\vec{\Lambda}|
\\
\\
 \tan^2\theta_{12}\sim \frac{\epsilon_1^2}{\epsilon_2^2} \qquad ;
 &\tan^2\theta_{13}\approx \frac{\Lambda_1^2}{\Lambda_2^2+\Lambda_3^2} 
\\
\\ 
 \tan^2\theta_{23}\approx  \frac{\Lambda_2^2}{\Lambda_3^2}
 \end{array}
\label{nuap}
\end{equation}
where we defined $\Lambda_i=\epsilon_iv_d+\mu v_i$ and we denoted
$|\vec{\Lambda}|=\sqrt{\Lambda_1^2+\Lambda_2^2+\Lambda_3^2}$ and
similarly for $|\vec{\epsilon}|$.

It is convenient to trade the soft parameters $B_i$ by $\Lambda_i$
since these are more directly related to the neutrino--neutralino mass
matrix. With this choice a point in the BRpV--mSUGRA parameter space is
specified by
\begin{equation}
m_0\,,\, m_{1/2}\,,\, \tan\beta\,,\, {\mathrm{sign}}(\mu)\,,\, 
A_0 \,,\,  \epsilon_i \: {\mathrm{, and}}\,\, \Lambda_i\,.
\end{equation}
Given a mSUGRA point, that is $m_0\,,\, m_{1/2}\,,\, \tan\beta\,,\,
{\mathrm{sign}}(\mu)$ and $ A_0 \,$, the BRpV parameters $\epsilon_i$
and $\Lambda_i$ are strongly constrained by the available neutrino
physics data~\cite{Diaz:2004fu}.  Therefore, the neutralino decay
properties are rather well determined, usually varying by up to
$\simeq 50$\% when we consider solutions for $\epsilon_i$ and
$\Lambda_i$ compatible with the experimental neutrino mixings and
squared mass differences at $3\sigma$ level, that we considered to
be~\cite{Maltoni:2004ei},
\begin{equation}
 \begin{array}{ll}
  \Delta m_{21}^2=7.6^{+0.7}_{-0.5}\times10^{-5}\,\text{eV}^2 \qquad;
 &\Delta m_{31}^2=2.4\pm0.4\times10^{-3}\,\text{eV}^2\,,
\\
\\
  \tan^2\theta_{12}=0.47^{+0.20}_{-0.12} \qquad ;
 &\tan^2\theta_{23}=1.00^{+1}_{-0.48},
\\
\\
  \tan^2\theta_{13}<0.05
 \end{array}
\label{nupa}
\end{equation}

In order to determine the BRpV parameters compatible with the neutrino
data, we use an iterative procedure whose starting point is the
approximated analytic results given in Eq.~(\ref{nuap}).  In each
iteration of the search procedure, the neutrino masses and mixing are
obtained numerically and compared with their $3\sigma$ ranges given by
Eq.~(\ref{nupa}).  If $\Delta m^2 _{23}$ is less (greater) than the
experimental range, then the value of $|\vec{\Lambda}|$ is increased
(decreased). The same is done for $\Delta m_{12}^2$ with
$|\vec{\epsilon}|$. Once proper values for $|\vec{\Lambda}|$ and
$|\vec{\epsilon}|$ are obtained, the interactions continues by
changing the values for $\Lambda_i$ and $\epsilon_i$ until a solution
for the squared mass differences and mixing angles is obtained.
Due to small neutrino masses, the BRpV interactions compatible with
the neutrino data are rather weak, with $|\vec \epsilon|$
($|\vec\Lambda|$) being of the order of $10^{-2}$--$10^{-1}$ GeV
(GeV$^2$).

In parameter space regions where the lightest neutralino is the LSP
the BRpV interactions render the lightest neutralino unstable and can
mediate two-- and three--body LSP decays depending on the SUSY
spectrum.  Considering the new BRpV interactions the lightest
neutralino can exhibit fully leptonic decays

\begin{itemize}

\item[$\star$] $\tilde{\chi}^0_1 \to \nu \ell^+ \ell^-$ with $\ell =e$
  or $\mu$;

   \item[$\star$] $\tilde{\chi}^0_1 \to \nu \tau^+ \tau^-$;

   \item[$\star$]  $\tilde{\chi}^0_1 \to \nu \tau^\pm \ell^\mp$;

\end{itemize}
as well as semi-leptonic decay modes 
\begin{itemize}

  \item[$\star$] $\tilde{\chi}^0_1 \to \nu q \bar{q}$;

  \item[$\star$] $\tilde{\chi}^0_1 \to \tau q^\prime \bar{q}$;

  \item[$\star$] $\tilde{\chi}^0_1 \to \ell q^\prime \bar{q}$;
    
  \item[$\star$] $\tilde{\chi}^0_1 \to \nu b \bar{b}$.

\end{itemize}
Notice, if kinematically allowed, some of these modes are generated in
two steps. Initially we have a neutralino two--body decay, like
$\tilde{\chi}^0_1 \to W^\mp \mu^\pm$, $\tilde{\chi}^0_1 \to W^\mp
\tau^\pm$, $\tilde{\chi}^0_1 \to Z \nu$, or $\tilde{\chi}^0_1 \to h
\nu$, followed by the $Z$, $W^\pm$ or $h$ decay; for further details
see Ref.~\cite{deCampos:2007bn}.  In addition to these channels there
is also the possibility of the neutralino decaying invisibly into
three neutrinos, however, this channel does not dominate in most of
the parameter space.
 
\begin{figure}[!h]
 \begin{center}
        \includegraphics[width=15cm]{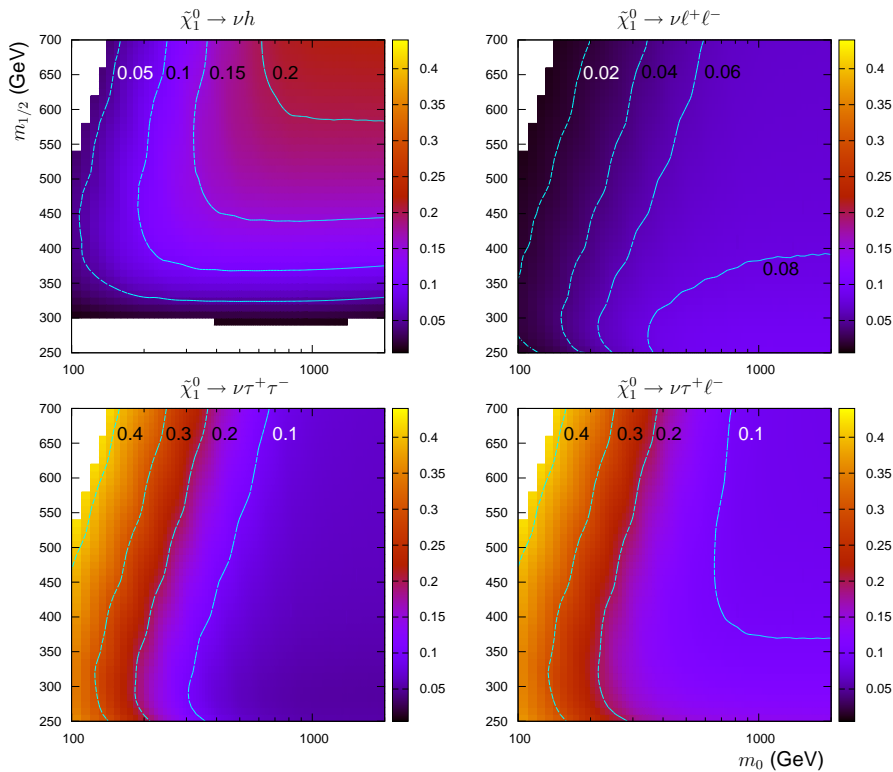}
 \end{center}
        \caption{Lightest neutralino branching ratios as a function of
          $m_0$ and $m_{1/2}$ for $A_0=-100$ GeV, $\tan\beta=10$, and
          $\mu>0$. The upper left (right) panel presents the branching
          ratio into $\nu h$ ($\nu\ell^+\ell^-$) while the lower left
          (right) panel is for $\nu\tau^+\tau^-$
          ($\nu\ell^\pm\tau^\mp$).}
\label{fig:brlep}
\end{figure}

\begin{figure}[!h]
 \begin{center}
        \includegraphics[width=15cm]{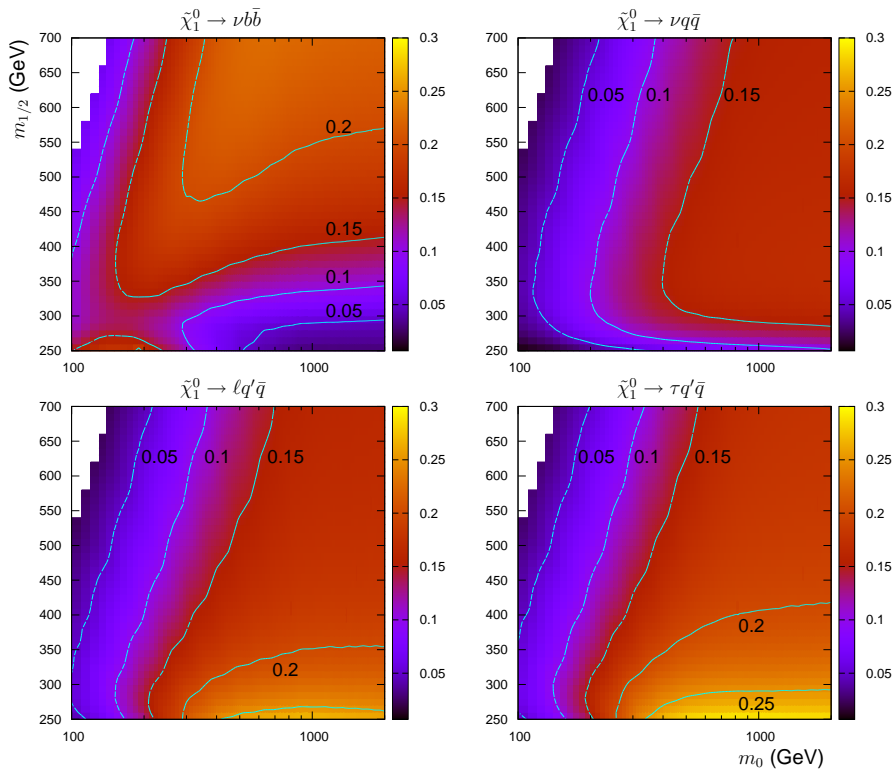}
 \end{center}
        \caption{Same as Figure \ref{fig:brlep} for the decays $\nu b
          \bar{b}$ (upper left panel), $\nu q \bar{q}$ (upper right
          panel), $\ell q^\prime \bar{q}$ (lower right panel), and
          $\tau q^\prime \bar{q}$ (lower right panel).  }
\label{fig:brsem}
\end{figure}

We depicted in Figures \ref{fig:brlep} and \ref{fig:brsem} the main
lightest neutralino branching ratios a function of $m_0 \otimes
m_{1/2}$ for $A_0=-100$ GeV, $\tan\beta=10$, and $\mu > 0$.  In the
white region in the top left corner of this figure the lightest
neutralino is no longer the LSP, that turns out to be the lightest
charged scalar ($S_1^\pm$). In this region the $\tilde{\chi}^0_1$
decays promptly into $S_1^\pm \tau^\mp$. Moreover, the $S_1^\pm$
decays so rapidly through BRpV interactions that it does not yield a
displaced vertex. In the small $m_0$ where the lightest neutralino is
the LSP and this decay is kinematically forbidden, the almost on-shell
$S_0^1$ contribution dominates and the main LSP decay modes are
$\nu\tau^+\tau^-$ and $\nu \tau^\pm \ell^\mp$; see the lower panels of
Fig.~\ref{fig:brlep}.


We can see from the top right panel of
Fig~\ref{fig:brlep} that $\tilde{\chi}_1^0 $ does have a sizeable
branching ratio into $\nu h$, specially for heavier LSP and larger
$m_0$. Notice that this decay is kinematically forbidden in the white
region in the bottom of this panel. From the other panels of this
figure we can learn that the leptonic decay $\nu \ell^+ \ell^-$ with
$\ell^\pm=\mu^\pm$ is of the order of a few to 10\%, while the decay
modes $e^\pm$, $\nu \tau^+ \tau^-$ and $\nu \tau^\pm \ell^\mp$ vary
from $\approx 40$\% at small $m_0$ to a few percent at large $m_0$. At
moderate and large $m_0$, these decays originate from the lightest
neutralino decaying into the two--body modes $\tau^\pm W^\mp$,
$\mu^\pm W^\mp$ and $\nu Z$, followed by the leptonic decay of the weak
gauge bosons.

\begin{figure}[t]
  \begin{center}
  \includegraphics[width=6.5cm]{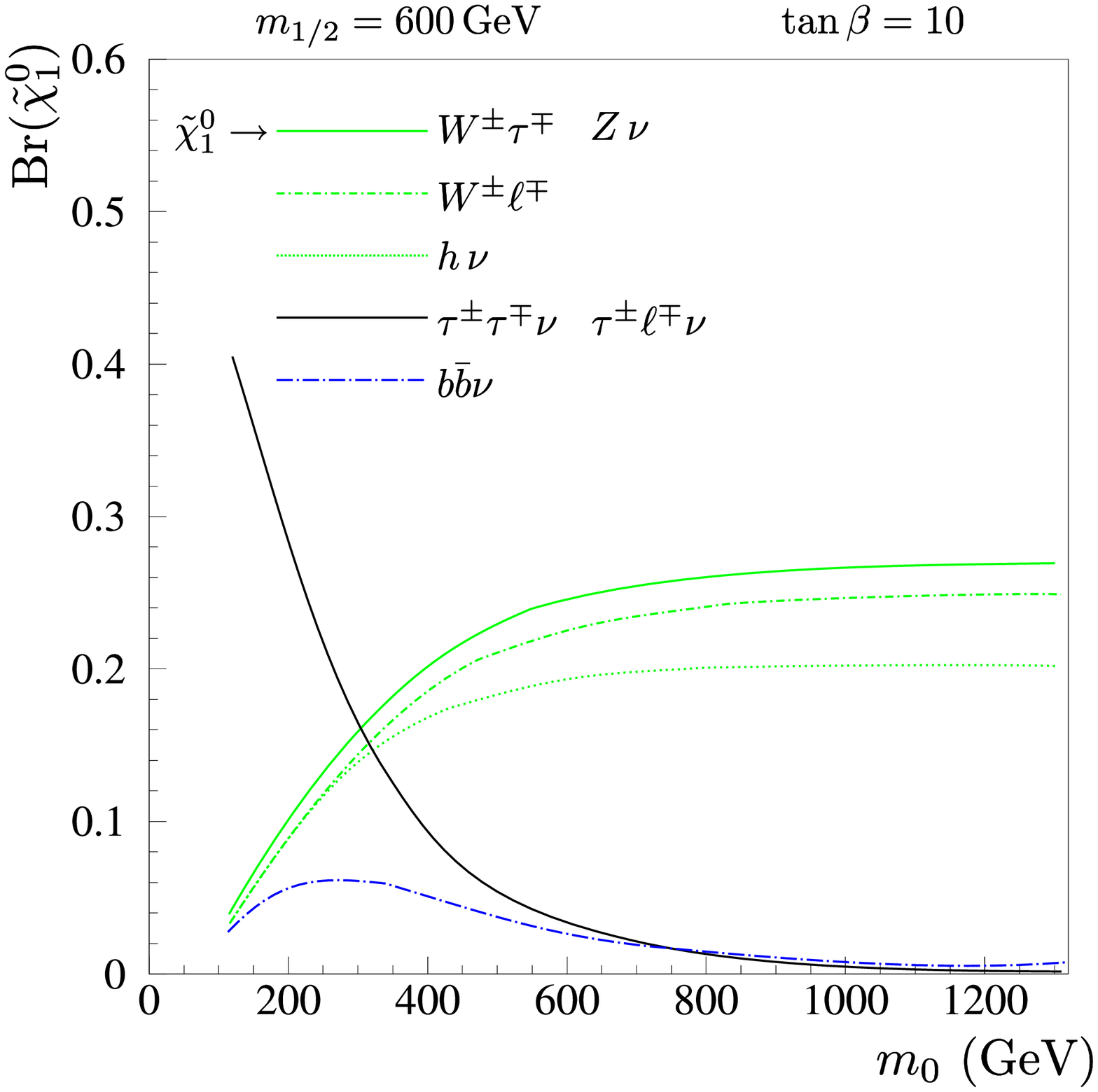}
  \includegraphics[width=6.5cm]{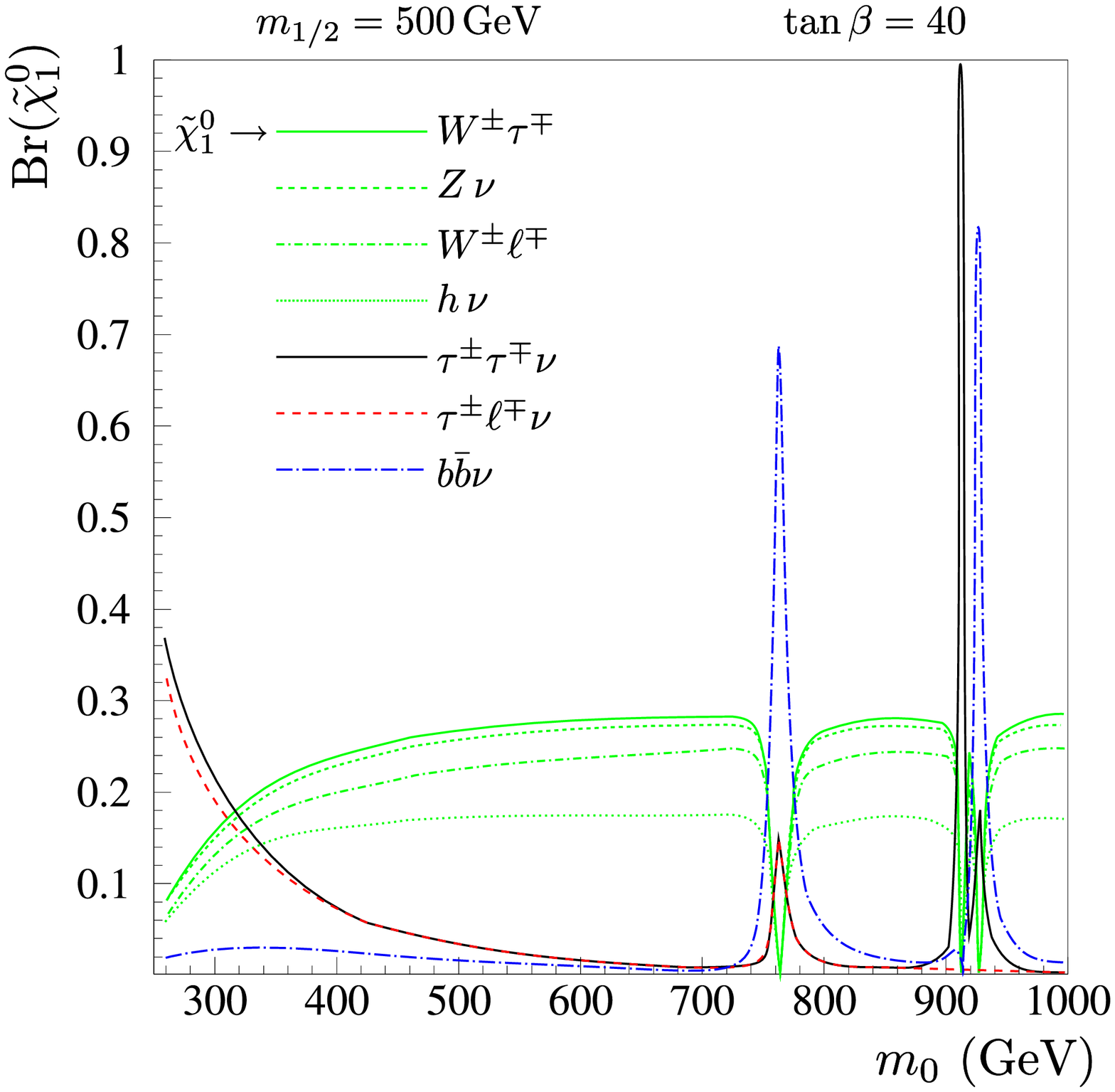}
  \end{center}
  \vspace*{-8mm}
  \caption{ $\tilde\chi_1^0$ branching ratios versus $m_0$ for $A_0=-100$
    GeV, $\mu > 0$, $m_{1/2}= 500$ GeV, and $\tan\beta=10$ (left panel)
 or$\tan\beta=40$ (right panel).  }
\label{fig:br40}
\end{figure}

In general, semi-leptonic decays of the LSP are suppressed at small
$m_0$ and they dominate at large $m_0$ due to two--body decays; see
Fig.~\ref{fig:brsem}.  In fact, a closer look into this figure
reveals that the dominant decay modes in this $m_0$ region are $
\tau^\mp W^\pm \to \tau^\mp q^\prime \bar{q}$ , $ \ell^\mp W^\pm \to
\ell^\mp q^\prime \bar{q}$, $\nu Z\to \nu q \bar{q}$, and $\nu h \to
\nu b \bar{b}$. For a given value of $m_{1/2}$, the branching ratios
into $\tau^\mp W^\pm $, $\nu Z $ and $\ell^\mp W^\pm $ are dominant
and rather similar. Moreover, the importance of the $\nu h$ channel
grows as $m_{1/2}$ increases in the moderate and large $m_0$ regions;
these facts are illustrated in the right panel of Fig.~\ref{fig:br40}
for a fixed value of $m_{1/2}$.


\begin{figure}[t]
  \begin{center}

  \includegraphics[width=6.5cm]{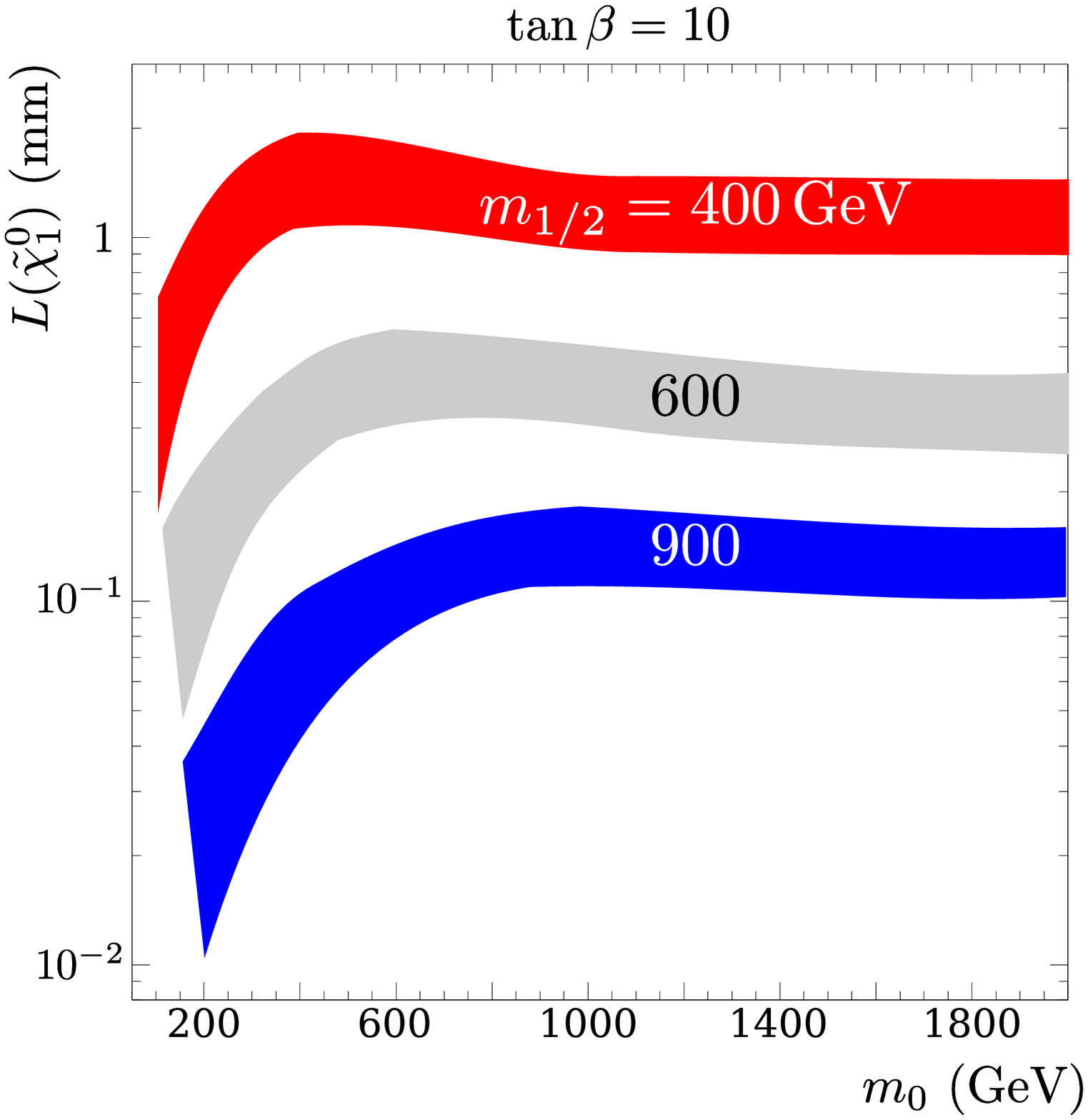}
  \includegraphics[width=6.5cm]{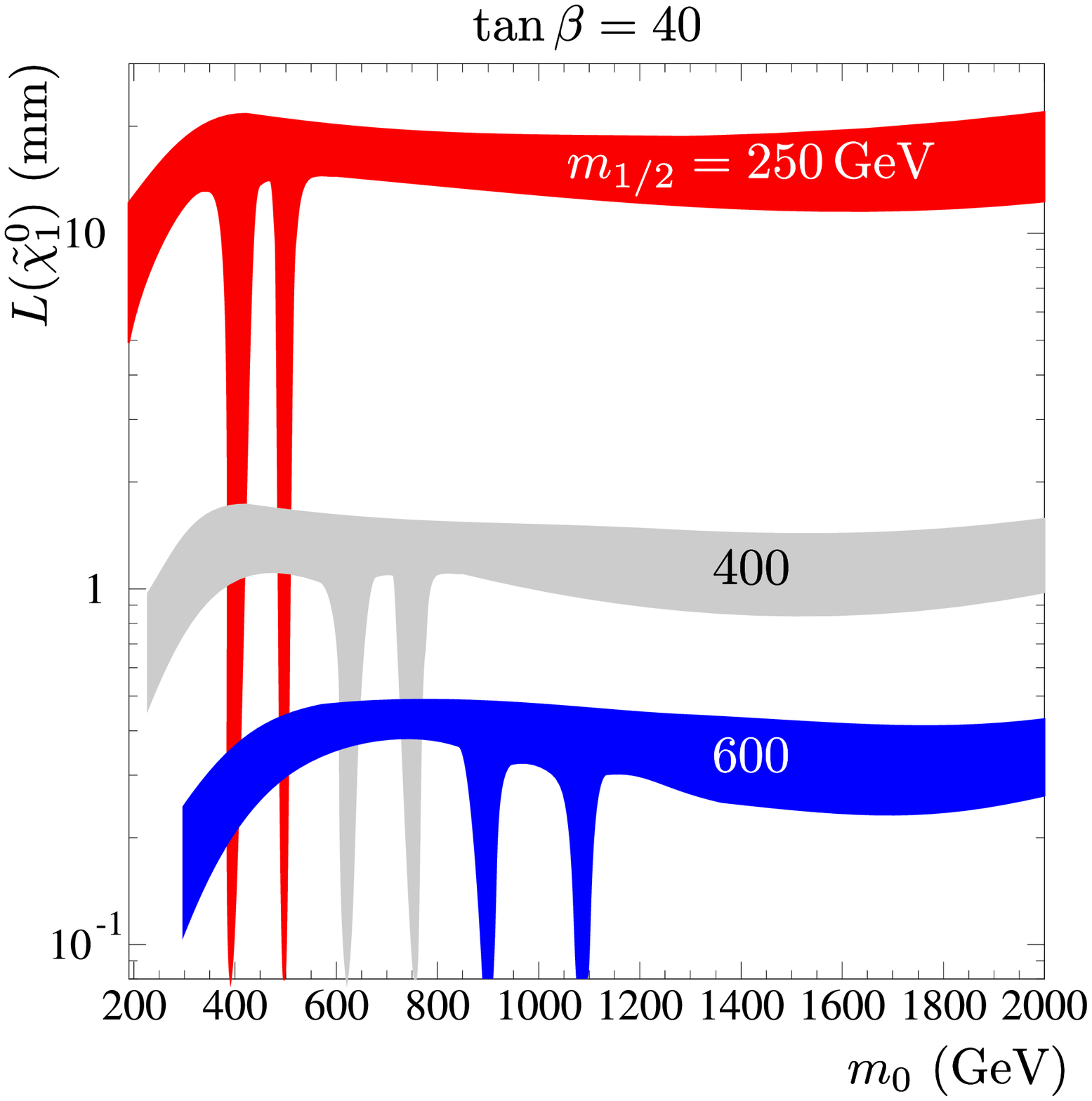}
  \end{center}
  \vspace*{-8mm}
  \caption{ $\tilde\chi_1^0$ decay length versus $m_0$ for $A_0=-100$
    GeV, $\mu > 0$, several values of $m_{1/2}$, and $\tan\beta=10$
    (left panel) and $\tan\beta=40$ (right panel).  The widths of the
    three shaded (colored) bands for fixed $m_{1/2}$ GeV correspond to
    the variation of the BRpV parameters in such a way that the
    neutrino masses and mixing angles fit the required
    values within $3\sigma$~\cite{Maltoni:2004ei}. }
\label{fig:ldec}
\end{figure}


The BRpV--mSUGRA parameters needed to reproduce the present neutrino
data are rather small ~\cite{Diaz:2003as,Hirsch:2000ef}. Therefore,
the BRpV interactions responsible for the LSP decay are feeble,
rendering the LSP long lived as can be seen from Figure
\ref{fig:ldec}. The left panel of this figure was obtained for
$\tan\beta=10$ and it shows that the $\tilde{\chi}^0_1$ decay length
can be as large as a few millimeters for light neutralino masses. As
the $\tilde{\chi}^0_1$ mass ($m_{1/2}$) increases its decay length
shortens, however, it is still sizeable even for heavier neutralinos
($\simeq 100~\mu$m). Notice that for fixed values of $m_{1/2}$ the
decay length is rather independent of $m_0$.  Another salient feature
of the BRpV--mSUGRA model is that the neutralino decay length is
rather insensitive to variations of the BRpV parameters provided they
lead to neutrino masses and mixings compatible within $3\sigma$ with
the allowed experimental results. We present in the left panel of
Fig.~\ref{fig:ldec} the effect of such variations of the BRpV
parameters as bands for a fixed value of $m_{1/2}$.

In order to assess the dependence of our results on $\tan\beta$ we
present in the right panel of Fig.~\ref{fig:ldec} the lightest
neutralino decay length as a function of $m_0$ for $\tan\beta=40$.  As
we can see, the $\tan\beta=40$ results are rather similar to the ones
for $\tan\beta=10$, except for the appearance of narrow regions where
the $\tilde{\chi}^0_1$ decay length is drastically reduced. In these
special regions the mixing between scalars (higgses and sfermions) is
rather large which enhances some two--body decay channels. This can be
seen more clearly by comparing the two panels of Fig.~\ref{fig:br40}
where we present the LSP branching ratios as a function of $m_0$ for
$\tan\beta=10$ (left panel) and $\tan\beta=40$ (right panel). 
The $\nu b \bar{b}$ ($\nu h$) has a sudden increase around $m_0 \simeq
750$ and 950 GeV due to the large sneutrino--neutral higgs mixing,
while a large stau--charged Higgs mixing leads to a large increase of
the $\nu \tau^\pm \tau^\mp$ mode around $m_0 \simeq 900$ GeV. In
general these mixings are small since they are proportional to the
square of the BRpV parameters divided by the difference of the squared
MSSM masses~\cite{DeCampos:2001wq}, however, the mixings become quite
large in the regions presenting almost degenerate states. Therefore,
we can infer from Figure \ref{fig:br40} that the general trend
observed for $\tan\beta=10$ is maintained for $\tan\beta=40$, except
for the parameter space regions where sfermions and higgses are nearly
degenerate.


The existence of long decay lengths is an important feature of the
BRpV models since detached vertices are a smoking gun of SUSY with
R--parity violation.  Furthermore, displaced vertices provide an
additional handle that makes possible the SUSY search at the LHCb
experiment.

\section{Results and conclusions}
\label{results}

Our goal is to assess the potential of the LHCb experiment to search
for displaced vertices associated to the visible LSP decays.  In order
to generate the mass spectrum, widths and branching ratios in the
BRpV--mSUGRA model we employed a generalized version of the SPHENO
program~\cite{Porod:2003um}.  Given the mSUGRA parameters ($m_0\,,\,
m_{1/2}\,,\, \tan\beta\,,\, {\mathrm{sign}}(\mu)\,,\, A_0$) SPHENO
obtains a set of BRpV parameters that is compatible with the neutrino
data, as well as, the masses and branching ratios, writing its output
in the SLHA format~\cite{Skands:2003cj}. We carried out the event
generation using PYTHIA version
6.408~\cite{Sjostrand:2000wi,Sjostrand:1993yb} inputting the SLHA
SPHENO output. Since the BRpV parameters consistent with neutrino data
are rather small, the production takes place through the usual
R--parity conserving channels contained in PYTHIA. For the same
reason, the decay of all supersymmetric particles are dominated by the
R--conserving modes, except for the cases that it is suppressed or
forbidden, for instance, the LSP decays via R--parity violating
channels.

In our analysis we looked for one detached vertex away from the primary
vertex requiring it to be outside an ellipsoid around the primary vertex
\[
      \left ( \frac{x}{\delta_{xy}} \right )^2
   +  \left ( \frac{y}{\delta_{xy}} \right )^2
   +  \left ( \frac{z}{\delta_{z}} \right )^2   = 1 \; ,
\] 
where the $z$-axis is along the beam direction and $\delta_{xy} =
20~\mu$m and $\delta_z = 500~\mu$m. We further required that the LSP
decay vertex must be inside the vertex locator sub-detector; this is a
conservative requirement that can be traded by the existence of a
reconstructed detached vertex in a more detailed analysis. We took
into account the LHCb acceptance by considering in our study only the
LSP decay products in the $1.8 < \eta < 4.9$ pseudo--rapidity range.

Within the SM, detached vertices are associated to rather long lived
particles, like $B$'s and $\tau$'s. Therefore, the Standard Model
physics background can be eliminated by requiring that the tracks
associated to the displaced vertex give rise to an invariant mass
larger than 20 GeV. After imposing this cut, the neutralino displaced
vertex signal is background free. Consequently, we defined that a
point in the BRpV--mSUGRA parameter space is observable if 5 or more
events present a detached vertex passing the above requirements for a
given integrated luminosity.

Taking into account that the LHCb experiment already has a dimuon
trigger, we started our analysis looking for displaced vertices
associated to $\tilde{\chi}^0_1 \to \nu \mu^+ \mu^-$. Despite the
smallness of this branching ratio this is a very clean signal. In fact
the branching ratio into this decay mode varies from a from few per
mil at small $m_0$ to a few percent at moderate and large $m_0$ where
it origin are the two--body decays $\mu^\pm W^\mp$ and $\nu Z$.


\begin{figure}[t]
  \begin{center}
  \includegraphics[width=7.9cm]{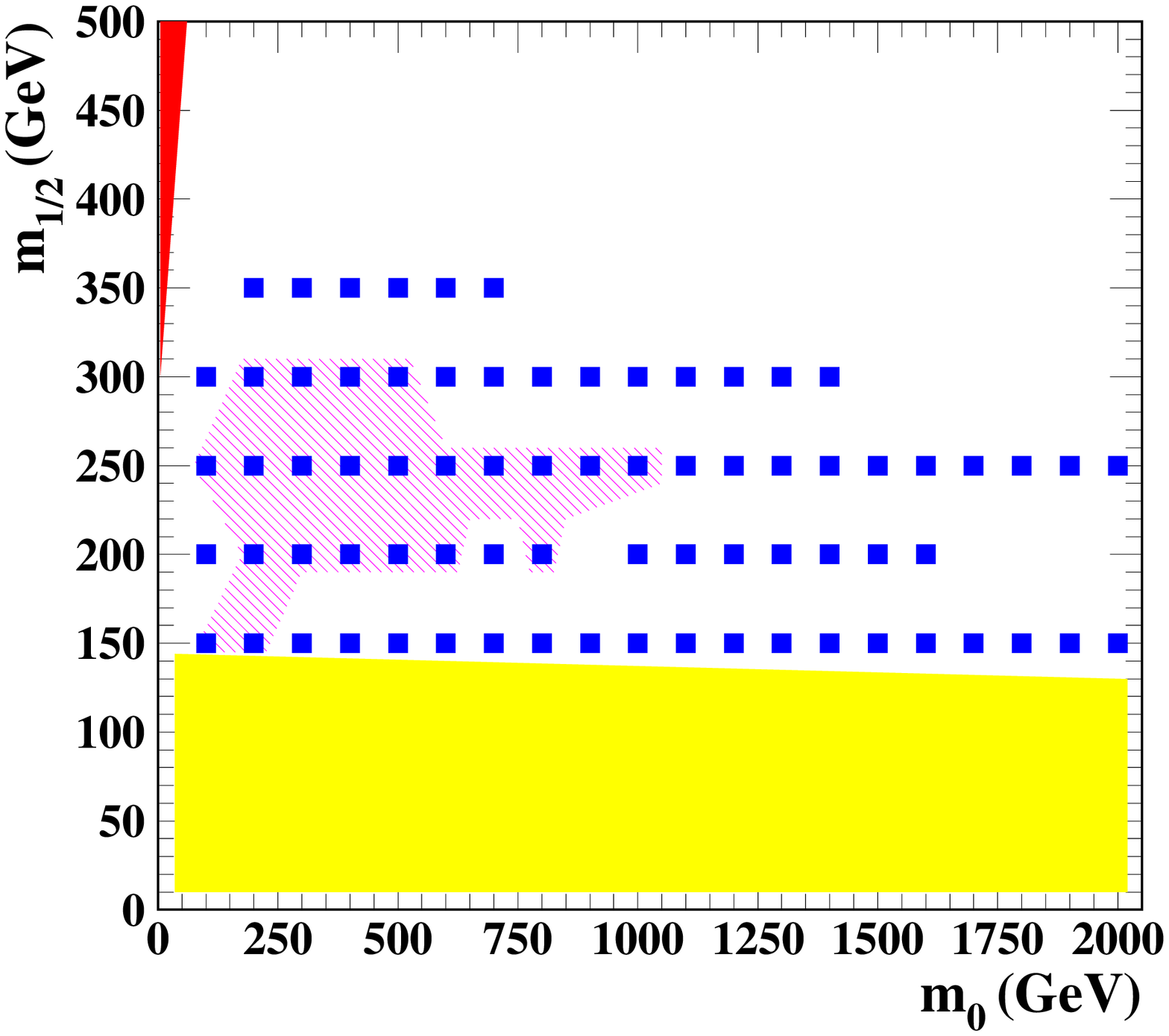}
  \includegraphics[width=7.9cm]{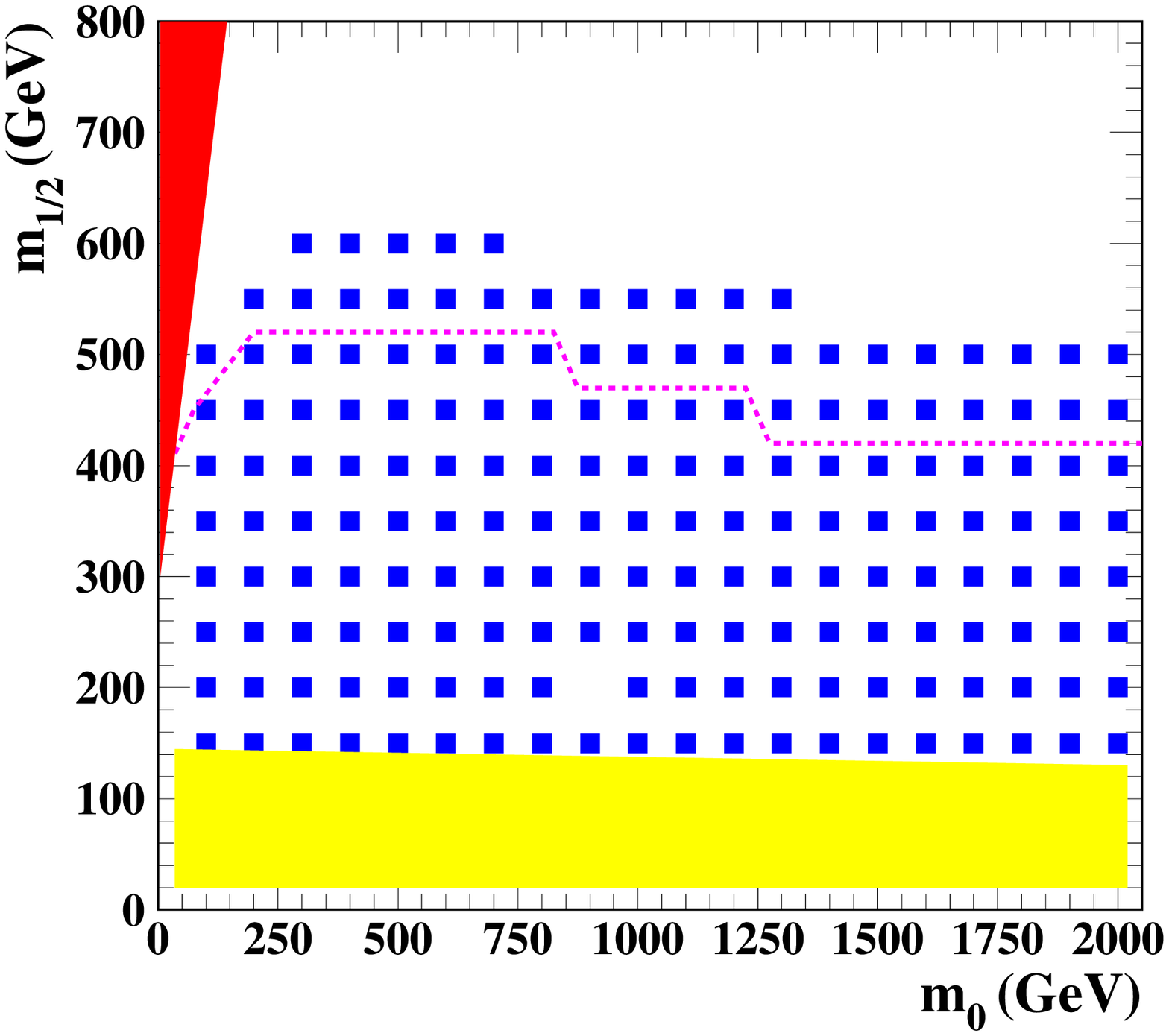}
  \includegraphics[width=7.9cm]{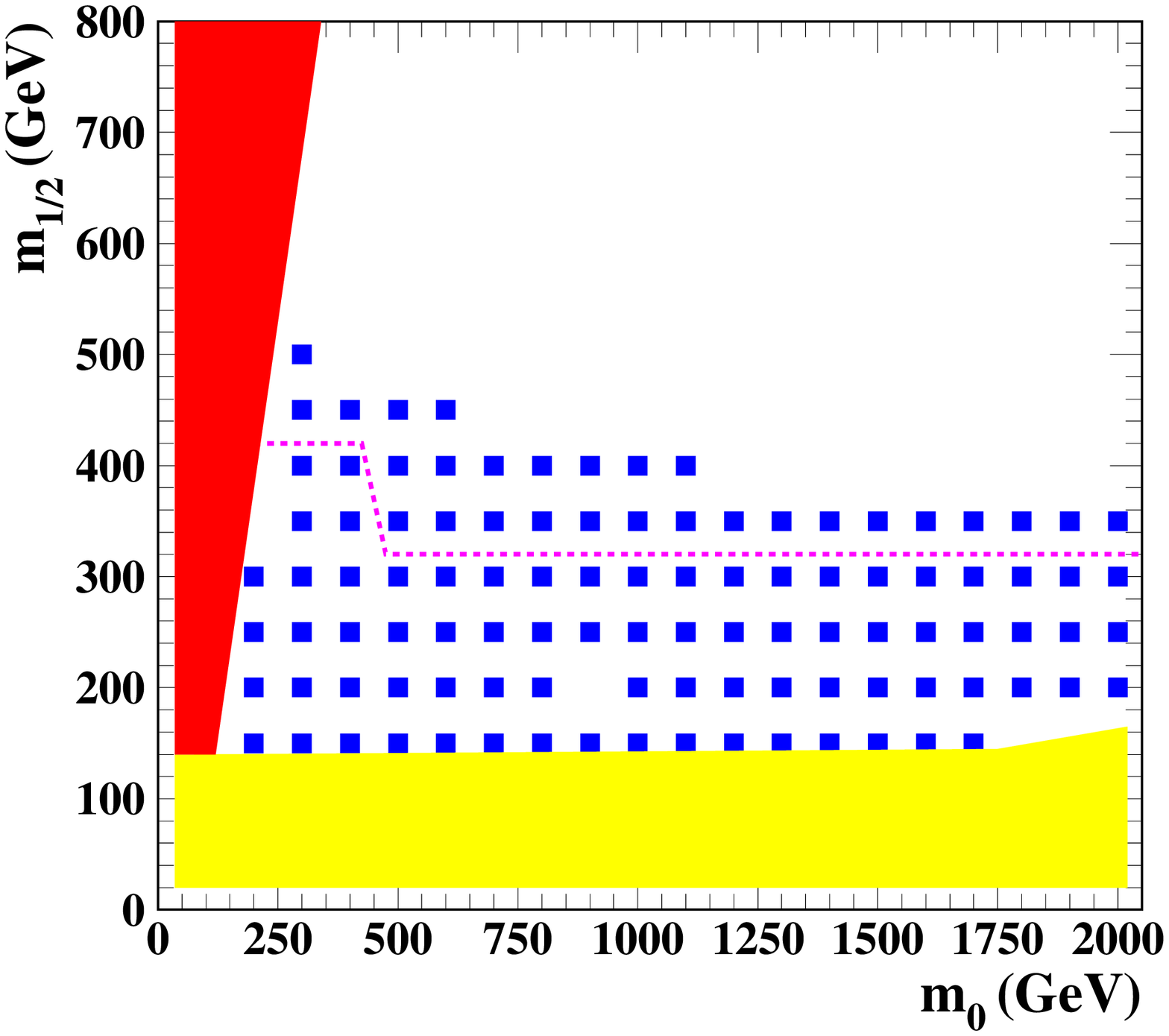}
  \includegraphics[width=7.9cm]{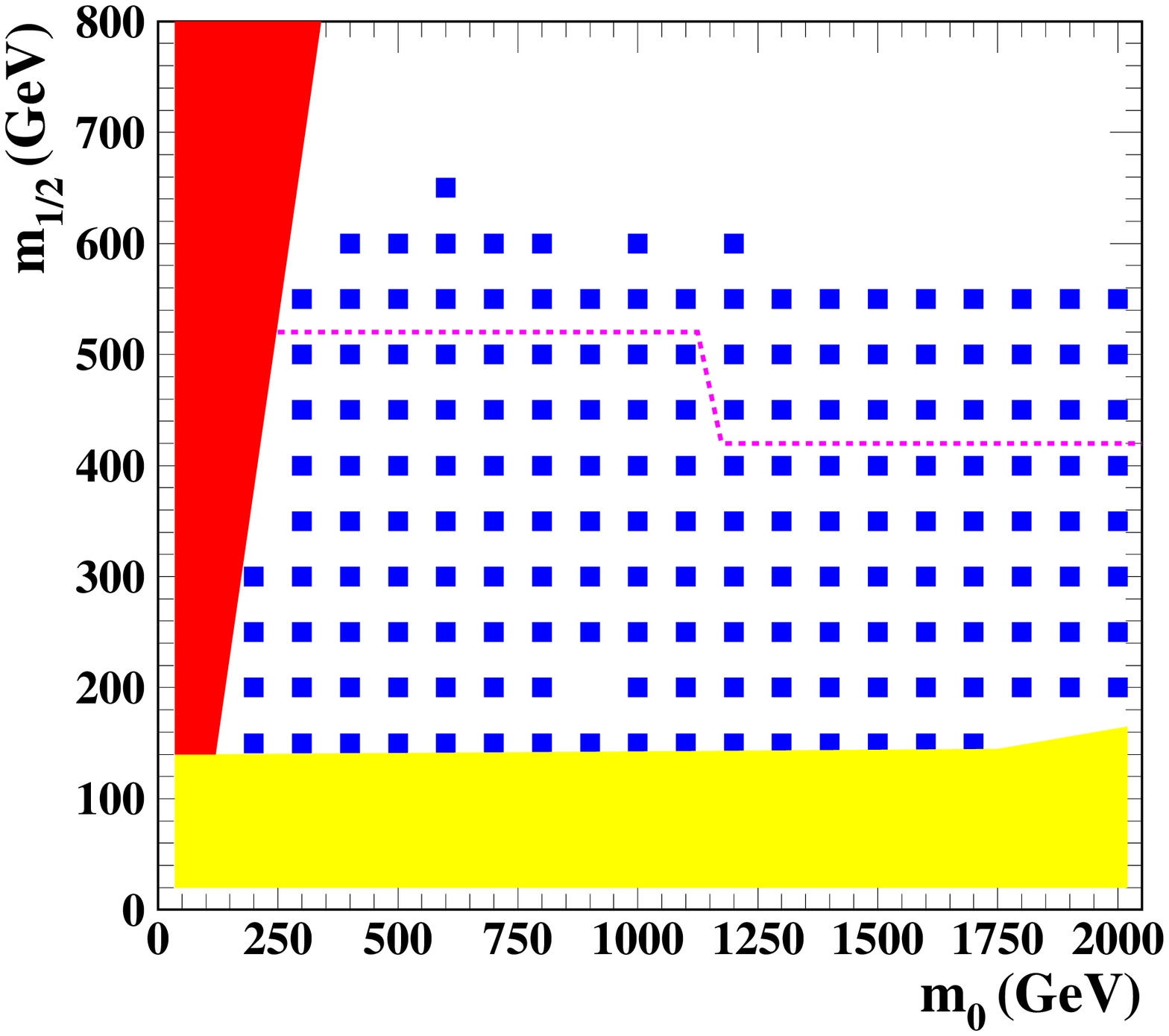}
  \end{center}
  \vspace*{-8mm}
  \caption{ The top left panel contains the LHCb discovery potential
    of BRpV--mSUGRA using dimuons coming from a detached vertex while
    the top right panel depicts the LHCb discovery potential using all
    visible modes.  Here we assumed that $A_0 = -100$ GeV,
    $\tan\beta=10$, and $\mu > 0$. The left (right) lower panel stands
    for the LHCb discovery potential using dimuons (all visible modes)
    for $\tan\beta=40$.  The blue squares stand for the points with 5
    or more events for an integrated luminosity of 2 fb$^{-1}$. The
    hatched area in the left panel and the region below the dashed
    line in the right panel correspond to a cross section greater than
    $10$ fb. The shaded region in the bottom is already excluded by
    the present available data, while the region in the top left
    corner of the figures indicates that the stau is the LSP. }
\label{fig:lhcb}
\end{figure}


We present in the top left panel of Figure~\ref{fig:lhcb} the region
of the $m_0 \otimes m_{1/2}$ plane where we expect 5 or more displaced
vertices for integrated luminosity of 2 fb$^{-1}$, $A_0=-100$ GeV,
$\tan\beta=10$, and $\mu>0$. As we can see, the LHCb can discover
dimuon displaced vertices for $m_{1/2}$ up to $\simeq 350$ (300) GeV
for small (moderate and large) values of $m_0$. These reaches
correspond to neutralino masses of the order of 120--140 GeV that
possess decay lengths of the order of a few millimeters.  The shaded
region in the top left corner of this panel of Fig.~\ref{fig:lhcb} is
the region where the stau is the lightest supersymmetric particle.
Since the stau has a very short lifetime in our BRpV--mSUGRA model, no
detached vertex signal is expected in this region. Furthermore, we
depicted as a hatched area the region where the production cross
section for dimuon detached vertices is greater than 10 fb, therefore,
leading to more than 20 background free events for an integrated
luminosity of 2 fb$^{-1}$.

It is interesting to notice that our BRpV--mSUGRA model predicts a
large number of additional purely leptonic decays of the lightest
neutralino, {\em e.g.} $\nu \tau^+\tau^-$ or $\nu \mu^\pm e^\mp$,
which possess a branching ratio similar or larger than the $\nu \mu^+
\mu^-$ mode. Moreover, the relative contribution of these channels is
determined by the neutrino physics~\cite{Hirsch:2003fe}.  Therefore,
it would be important to exam them, as well, to further test the
BRpV--mSUGRA model.

Since the LHCb experiment has very good vertex system we have also
studied its discovery potential using all LSP decays that allow 
vertex reconstruction, {\em i.e.} the decays exhibiting two or more
charged particles that can be reconstructed as emanating from the same
point.  We present in top right panel of Figure~\ref{fig:lhcb} the
region of the $m_0 \otimes m_{1/2}$ where 5 or more displaced vertices
can be reconstructed for the same choice of parameters used in the
dimuon analysis. As we can see, the inclusion of additional decay
modes substantially extends the LHCb search potential to $m_{1/2}
\simeq 600$ GeV at small $m_0$ and to $m_{1/2} \simeq 500$ GeV at
moderate and large $m_0$. Therefore, the search in all visible decay
mode allow us to explore neutralino masses up to 200--240 GeV and
decay lengths of a few tenths of a millimeter.  Notice that the LHCb
reach at moderate and large $m_0$ is approximately independent of this
parameter due to the dependence of the decay length
(Figure~\ref{fig:ldec}) and the branching ratios
(Figure~\ref{fig:brlep} and \ref{fig:brsem}) upon this parameter.
Moreover, for most of the parameter space where the signal is
observable the cross section is above $10$ fb, opening the possibility
of further detailed tests of the model in the event a signal is
observed.

As we have seen in the previous section, the LSP decay length is
rather stable against variations of the BRpV parameters provided they
are in $3\sigma$ agreement with the available neutrino
data. Therefore, the LHCb discovery reach does not change appreciably
when the BRpV parameters are varied, except for the boundary points of
the discovery region that can be shifted by a few GeV. In order to
assess the impact of $\tan\beta$ changes in the search for detached
vertices we present in lower panels of Figure~\ref{fig:lhcb} the reach
of the LHCb experiment for $\tan\beta=40$. As is well known for this
value of $\tan\beta$, the region where the stau is the LSP is larger
than in previous case.  Therefore, the large shaded region on the left
panel of this figure can not be explored via detached vertices since
the stau is short lived. Furthermore, we can see that the $m_{1/2}$
reach in the dimuon channel is increased by around $30$\% for small
and moderate $m_0$ ($\lesssim 1$ TeV), with the region exhibiting a
production cross section in excess of 10 fb being considerably
expanded. Notice that we have not indicated in this figure the small
non-observable regions where the neutralino decay length is
substantially reduced due to the presence of nearly degenerated
states.

The lower right panel of Fig.~\ref{fig:lhcb} depicts the area that the
detached vertex signal is visible when we consider all LSP decays that
allow vertex reconstruction for $\tan\beta=40$. In contrast with the
dimuon channel, the discovery area for 2 fb$^{-1}$ almost does not
change when we vary $\tan\beta$. This stability of the LHCb reach in
the visible modes could have been anticipated from the similar
behavior that the neutralino branching ratios and decay length have
for $\tan\beta = 10$ and 40, as seen in Figs.~\ref{fig:ldec} and
\ref{fig:br40}. Basically, the small changes in the branching ratios
introduced by the $\tan\beta$ variation is between visible decay
modes, consequently not affecting the overall signal reconstruction
efficiency.

It is interesting to contrast the LHCb discovery potential with the
ATLAS/CMS ones in the low luminosity initial run in order to estimate
the contribution to BRvP searches that the different experiments can
give. There are two main difference between the LHCb and ATLAS/CMS
experiments: first of all, the ATLAS/CMS will have a luminosity 5
times the LHCb one in this period. Secondly, ATLAS/CMS vertex
detectors have a larger pseudo-rapidity coverage, ranging from $-2.5 <
\eta < 2.5$.  Following Ref.~\cite{deCampos:2007bn}, we performed the
ATLAS/CMS analysis requiring the presence of two displaced vertices
that satisfy the above cuts, except for the fact that we considered
charged particles in the $|\eta| < 2.5$ pseudo-rapidity range.

\begin{figure}[t]
  \begin{center}
  \includegraphics[width=7.9cm]{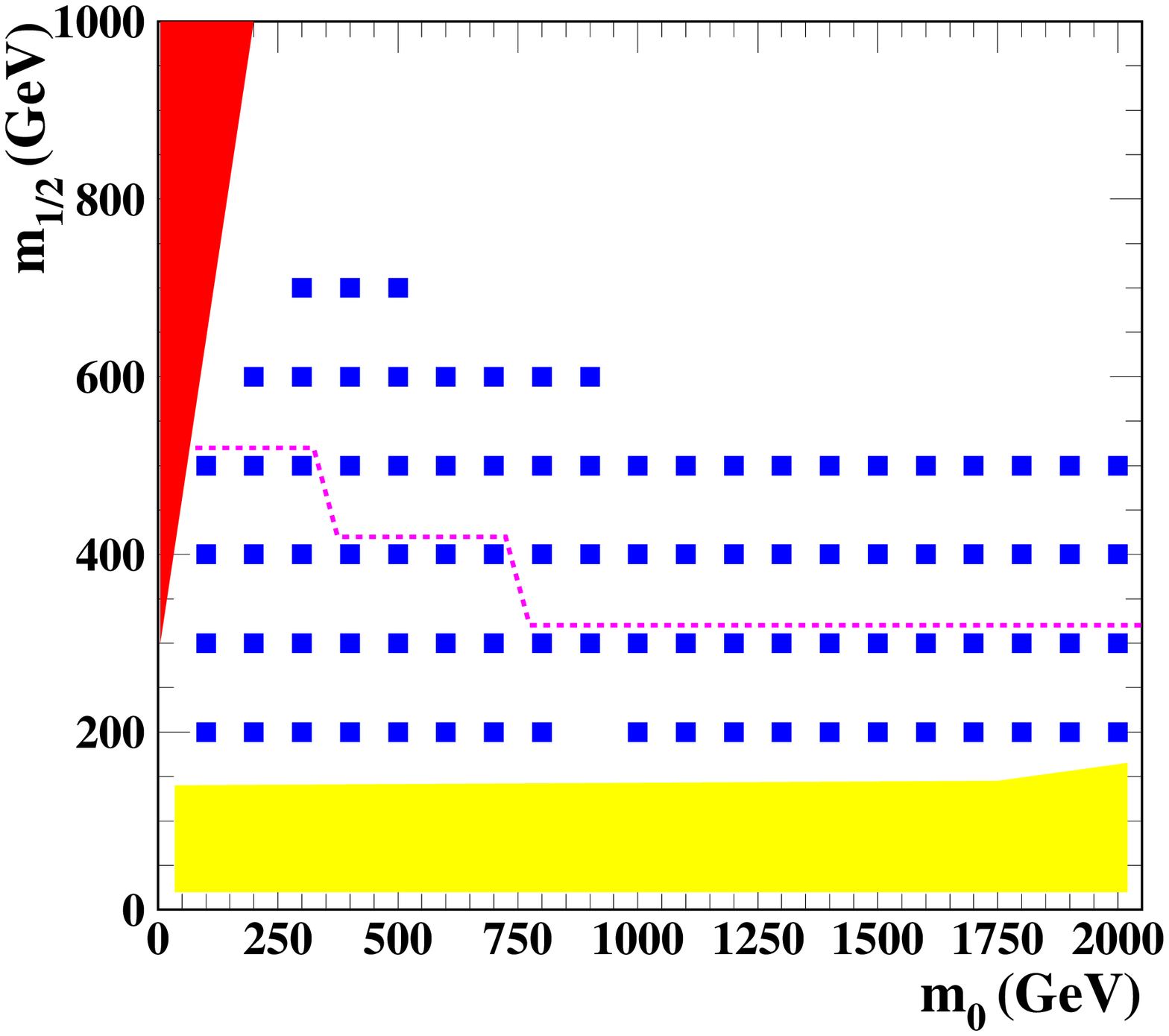}
  \includegraphics[width=7.9cm]{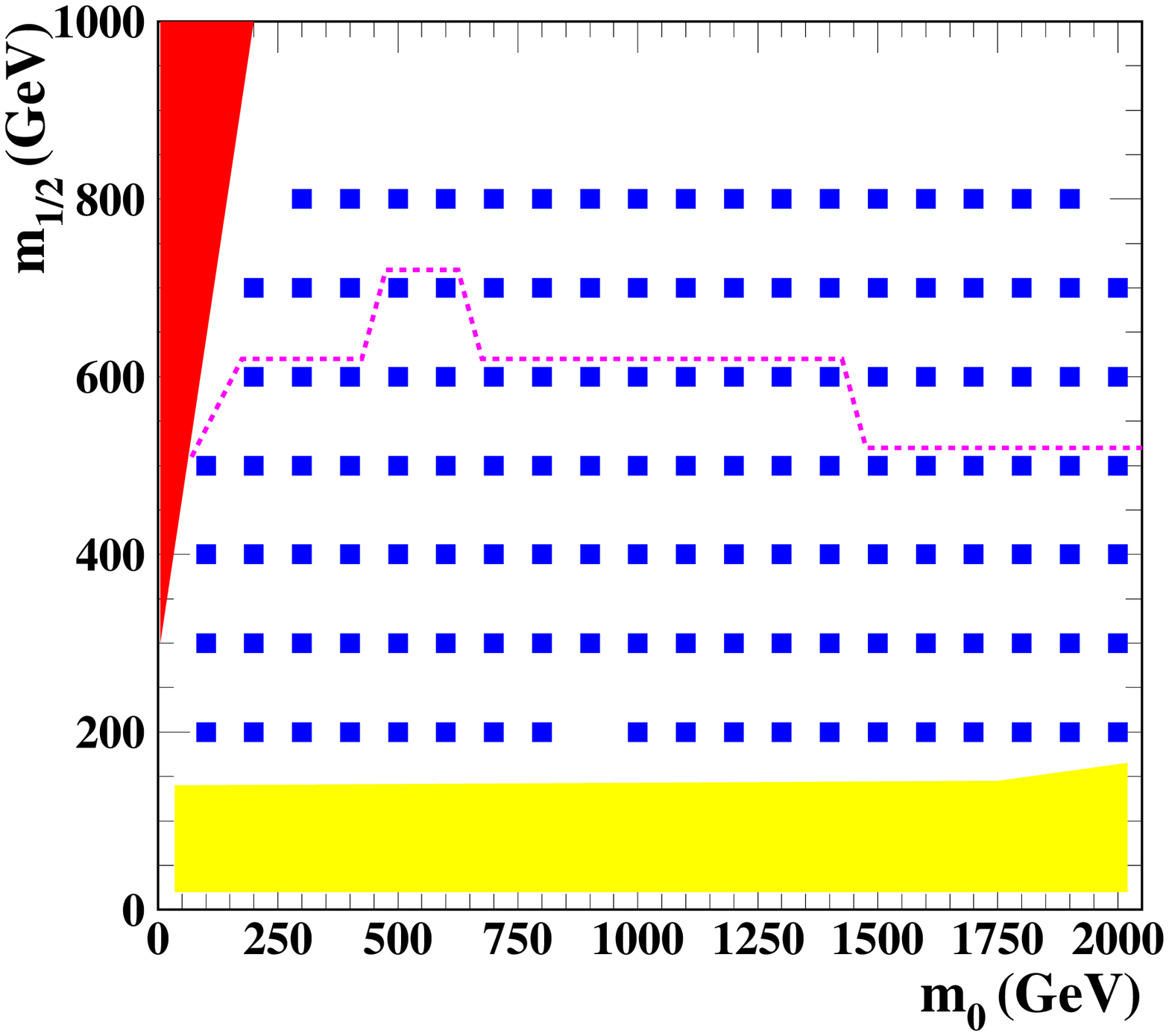}
  \includegraphics[width=7.9cm]{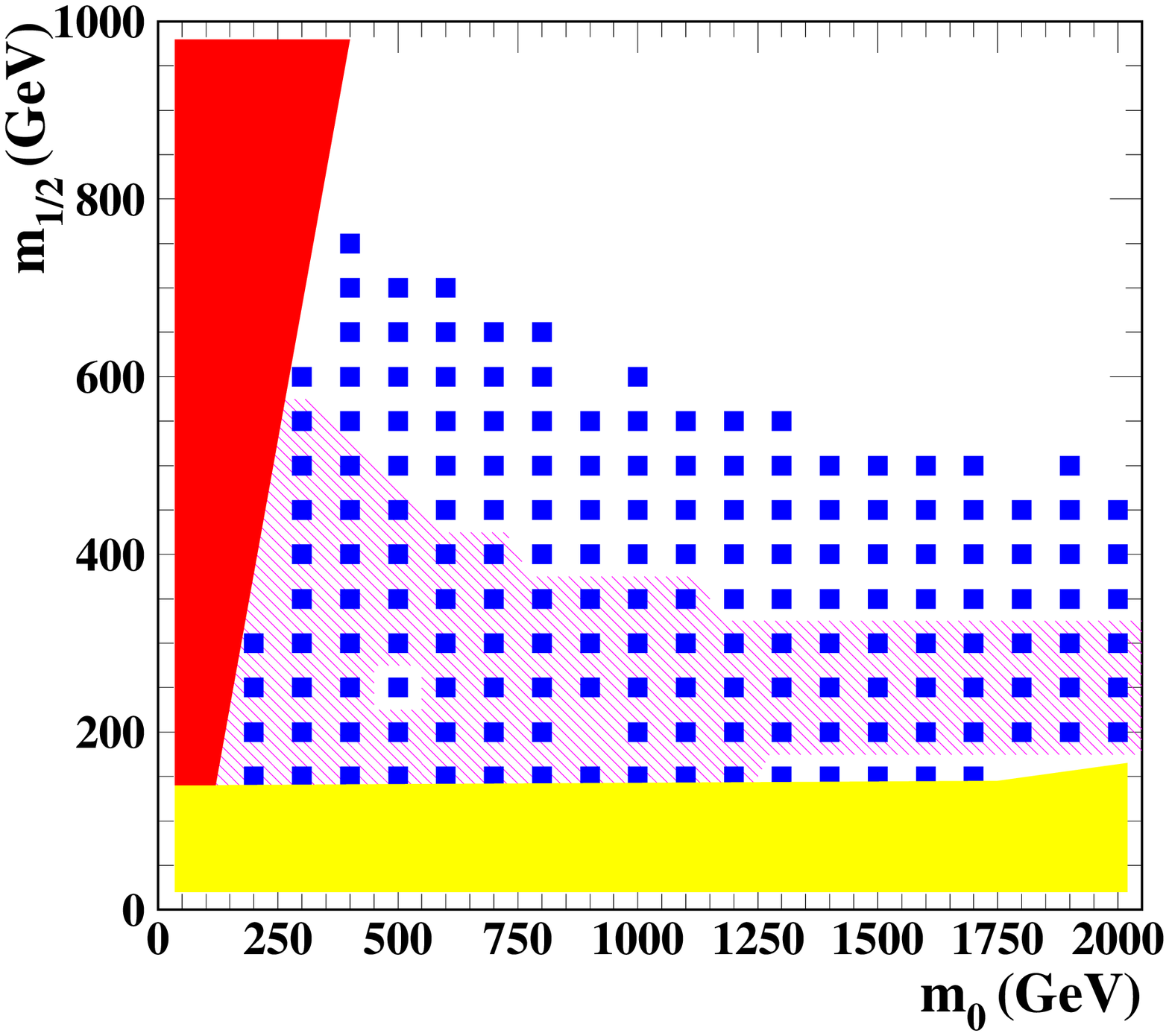}
  \includegraphics[width=7.9cm]{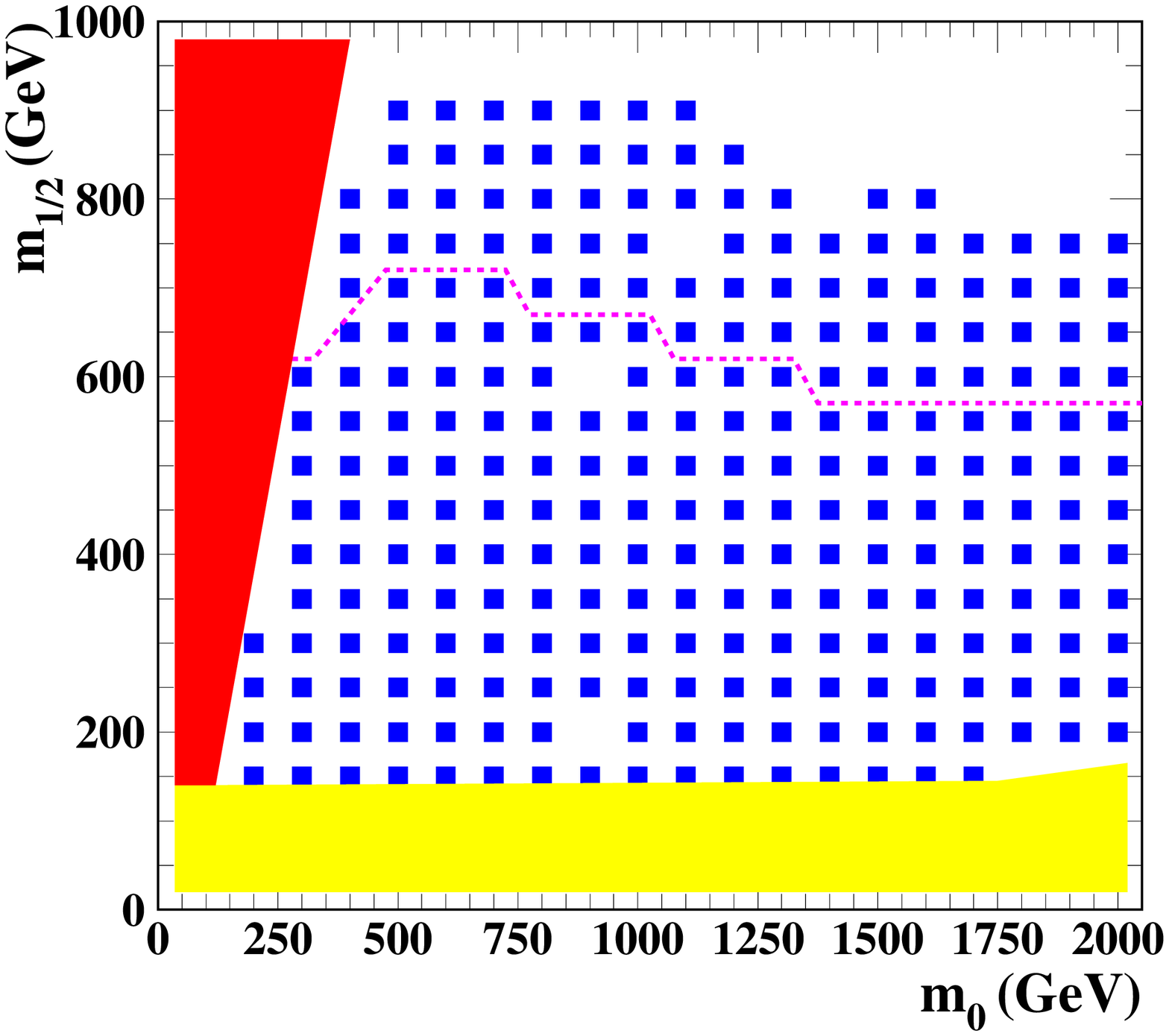}
  \end{center}
  \vspace*{-8mm}
  \caption{ Same as Figure~\ref{fig:lhcb} but for the ATLAS and CMS 
geometrical acceptance and an integrated luminosity of 10
fb$^{-1}$.}
\label{fig:atlas}
\end{figure}

We present in the top left panel of Fig.~\ref{fig:atlas} the ATLAS/CMS
reach in dimuon displaced vertices for the previous choice of
parameters, $\tan\beta=10$ and an integrated luminosity of 10
fb$^{-1}$.  From this figure we can see that ATLAS/CMS can probe
neutralino masses up to 290 GeV in dimuon displaced vertices.
Comparing this figure with the top left panel of Fig.~\ref{fig:lhcb}
we can see that ATLAS/CMS reach is a factor of 2 larger than the LHCb
reach in this channel. This larger ATLAS/CMS discovery potential
originates mainly from the larger planned integrated luminosity for
these detectors.


The results for the search of displaced vertices at ATLAS/CMS taking
into account all visible modes is presented in the top right panel of
Figure~\ref{fig:atlas} for $\tan\beta=10$.  As we can see form this
panel, adapted from Ref.~\cite{deCampos:2007bn}, ATLAS/CMS will be
able to probe $m_{1/2}$ up to 800 GeV which corresponds to neutralino
masses up to $\simeq 340$ GeV. Therefore, the LHCb reach in $m_{1/2}$
adding all sources of displaced vertices is around 70\% the
corresponding reach at ATLAS/CMS.  From the rapidity coverage of the
LHCb and ATLAS/CMS detectors, as well as, the prospects for the
integrated luminosities, one could expected that the discovery
potential of ATLAS/CMS would be considerably larger than the LHCb one.
Nevertheless, the reach in $m_{1/2}$ are rather similar for all
experiments due to the fast decrease of the SUSY production cross
section as $m_{1/2}$ increases.

The lower left (right) panel of Figure~\ref{fig:atlas} contains the
ATLAS/CMS discovery region in the dimuon (all visible) neutralino
decay mode. First of all, it is interesting to notice that the
ATLAS/CMS reach in the dimuon channel does not change significantly
as we vary $\tan\beta$. In the $\tan\beta = 40$ panel one can see the
effect of the drastic reduction of the neutralino decay length
observed previously in Fig.~\ref{fig:ldec}. For instance, the cross
section for the mSUGRA point
$m_{1/2} = 250$ GeV and $m_0 = 500$ GeV falls down under the 10 fb region
due to the reduction of the neutralino decay length. The same effect
can be seen in the lower right panel for the mSUGRA point $m_{1/2} =
600$ GeV and $m_0 = 900$ GeV, where the signal vanishes. 
Despite of this, the $m_{1/2}$
reach of these experiments is still much larger than the one expected
for the LHCb.  In the all visible neutralino decay mode, the ATLAS/CMS
discovery reach increases as $\tan\beta$ grows. Therefore, the LHCb 
$m_{1/2}$ reach in this channel is $\simeq 60$\% of the ATLAS/CMS
one. 

\begin{figure}[t]
  \begin{center}
  \includegraphics[width=7.9cm]{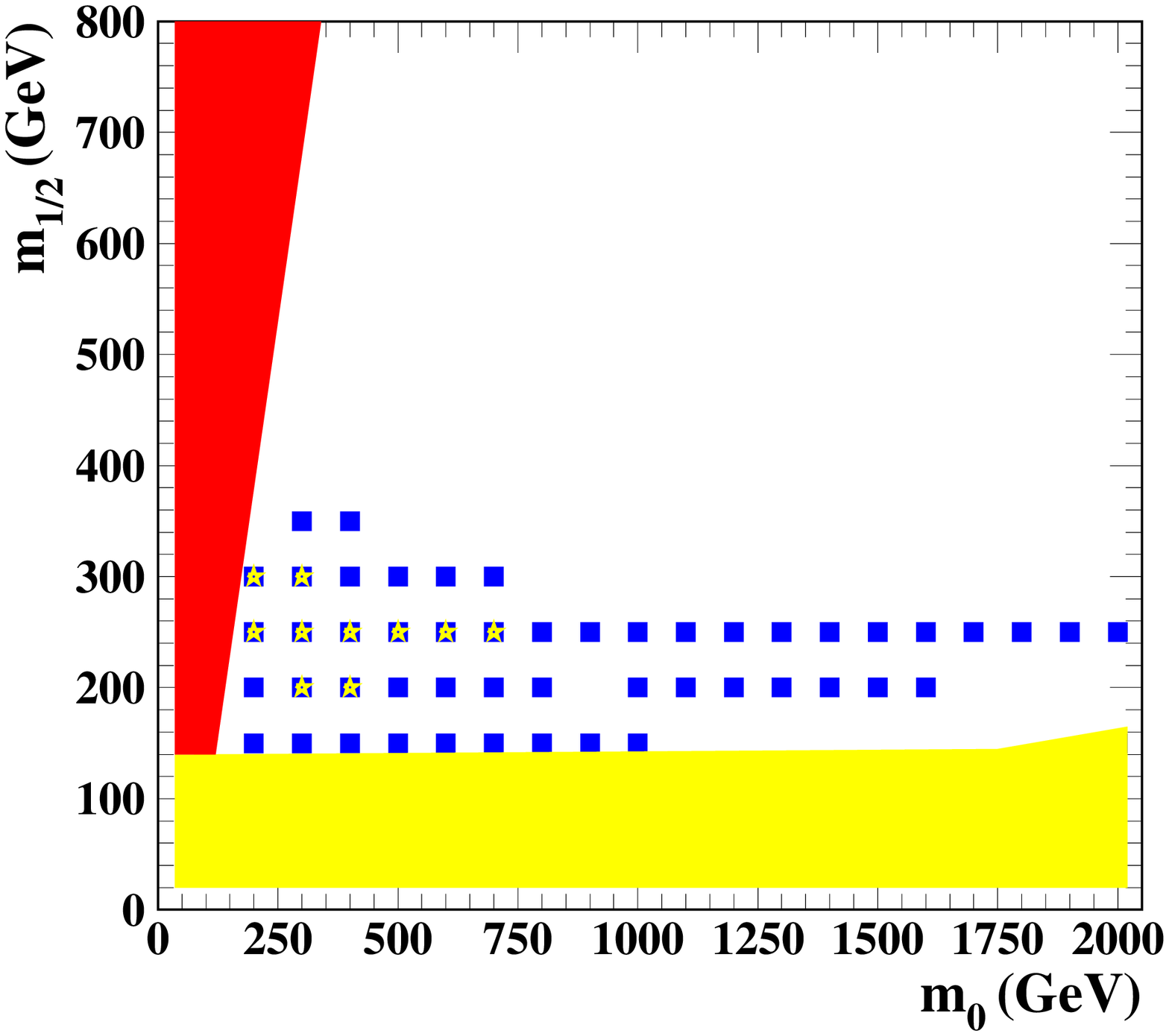}
  \includegraphics[width=7.9cm]{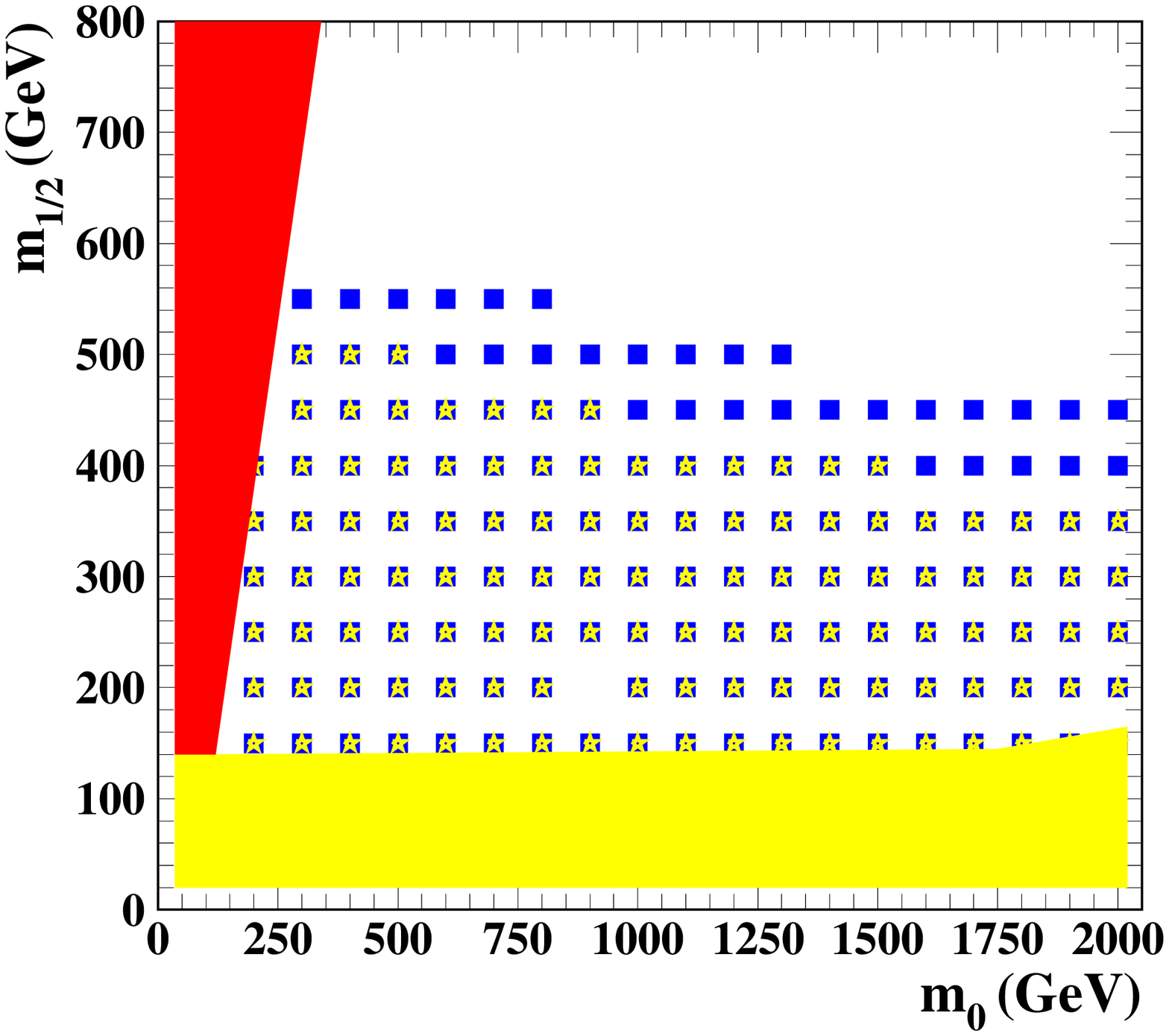}
  \end{center}
  \vspace*{-8mm}
  \caption{ Same as Figure~\ref{fig:lhcb} for $\tan\beta = 10$
    assuming a reconstruction  
    efficiency of 50\% and no background (blue squares) or 5
    background events (yellow stars).  }
\label{fig:pess}
\end{figure}

The results above should be taken with a grain of salt since our
analysis did not include the trigger and reconstruction efficiencies,
as well as instrumental backgrounds, which can only be estimated
passing the events through a full detector simulation. In order to
estimate the deterioration of the detached vertex signal we considered
a reconstruction and trigger efficiency of 50\% and that the number of
instrumental background events is either zero or five for an
integrated luminosity of 2 fb$^{-1}$.  Figure \ref{fig:pess} depicts
the LHCb discovery potential at the 5$\sigma$ level for these
scenarios assuming $A_0 = -100$ GeV, $\tan\beta = 10$, and $\mu >0$; the
background free scenario is represented by (blue) squares while the
existence of 5 background events case is given by the (yellow)
stars. From the left panel of Fig.\ \ref{fig:pess} the inclusion of
the reconstruction efficiency reduces the LHCb reach in the dimuon
channel at large $m_{1/2}$. On the other hand, if there were
additional background, the dimuon channel would be severely depleted.
The right panel of Fig.\ \ref{fig:pess} shows that the completely
inclusive search is depleted at large $m_0$ and $m_{1/2}$ in these
scenarios, however, not as much as in the dimuon topology.


\bigskip

\noindent{\bf Conclusions:}

\bigskip

Our results allow us to conclude that the LHCb can search for
supersymmetric models with bilinear R--parity breaking in a large
fraction of the parameter space due to the additional handle provided
by the existence of detached vertices in this class of models.
Moreover, the discovery potential of the LHCb is rather similar
(60--70\%) of the ATLAS/CMS one in the low luminosity run of the LHC.


\section*{Acknowledgments}

We thank J.\ W.\ F.\ Valle, L.\ de Paula, M.\ Gandelman and N.\
Gueissaz for enlightening discussions.  Work supported by Conselho
Nacional de Desenvolvimento Cient\'{\i}fico e Tecnol\'ogico (CNPq) and
by Funda\c{c}\~ao de Amparo \`a Pesquisa do Estado de S\~ao Paulo
(FAPESP); by Colciencias in Colombia under contract
1115-333-18740. M. B. Magro would like to acknowledge Instituto de
F\'{\i}sica-USP for hospitality.

\bibliographystyle{h-physrev}

\begin{thebibliography}{10}

\bibitem{Diaz:1997xc}
M.~A. D\'{\i}az, J.~C. Rom\~ao, and J.~W.~F. Valle,
\newblock Nucl. Phys. {\bf B524}, 23 (1998), [hep-ph/9706315].

\bibitem{chun:1998gp}
E.~J. Chun, S.~K. Kang, C.~W. Kim, and U.~W. Lee,
\newblock Nucl. Phys. {\bf B544}, 89 (1999);
D.~E. Kaplan and A.~E. Nelson,
\newblock JHEP {\bf 01}, 033 (2000);
F.~Takayama and M.~Yamaguchi,
\newblock Phys. Lett. {\bf B476}, 116 (2000);
T.~Banks, Y.~Grossman, E.~Nardi, and Y.~Nir,
\newblock Phys. Rev. {\bf D52}, 5319 (1995).

\bibitem{Diaz:2003as}
M.~A. D\'{\i}az {\em et~al.},
\newblock Phys. Rev. {\bf D68}, 013009 (2003), [hep-ph/0302021].

\bibitem{Hirsch:2000ef}
M.~Hirsch {\em et~al.},
\newblock Phys. Rev. {\bf D62}, 113008 (2000), [hep-ph/0004115],
\newblock Err-ibid. {\bf D65}:119901,2002.

\bibitem{DeCampos:2001wq} F.~de Campos, M.~A.~D\'{\i}az,
  O.~J.~P.~\'Eboli, M.~B.~Magro and P.~G.~Mercadante,
  Nucl.\ Phys.\ {\bf B623}, 47 (2002)
  [arXiv:hep-ph/0110049].

\bibitem{deCampos:2005ri}
F.~de~Campos {\em et~al.},
\newblock Phys. Rev. {\bf D71}, 075001 (2005), [hep-ph/0501153].

\bibitem{hall:1984id}
L.~J. Hall and M.~Suzuki,
\newblock Nucl. Phys. {\bf B231}, 419 (1984).

\bibitem{ross:1985yg}
G.~G. Ross and J.~W.~F. Valle,
\newblock Phys. Lett. {\bf B151}, 375 (1985).

\bibitem{ellis:1985gi}
J.~R. Ellis {\it et~al.},
\newblock Phys. Lett. {\bf B150}, 142 (1985).

\bibitem{marek:1996}
M.~Nowakowski and A. Pilaftsis (Rutherford),
\newblock Nucl. Phys. {\bf B461}, 19 (1996),
[hep-ph/9508271].


\bibitem{masiero:1990uj}
A.~Masiero and J.~W.~F. Valle,
\newblock Phys. Lett. {\bf B251}, 273 (1990);
J.~C. Rom\~ao, C.~A. Santos, and J.~W.~F. Valle,
\newblock Phys. Lett. {\bf B288}, 311 (1992);
J.~C. Rom\~ao, A.~Ioannissyan, and J.~W.~F. Valle,
\newblock Phys. Rev. {\bf D55}, 427 (1997), [hep-ph/9607401].

\bibitem{Hirsch:2004he}
M.~Hirsch and J.~W.~F. Valle,
\newblock New J. Phys. {\bf 6}, 76 (2004), [hep-ph/0405015].

\bibitem{deCampos:2007bn} F.~de Campos, O.~J.~P.~\'Eboli, M.~B.~Magro,
  W.~Porod, D.~Restrepo, M.~Hirsch and J.~W.~F.~Valle,
  JHEP {\bf 0805}, 048 (2008)
  [arXiv:0712.2156 [hep-ph]].


\bibitem{Magro:2003zb} M.~B.~Magro, F.~de Campos, O.~J.~P.~\'Eboli,
  W.~Porod, D.~Restrepo and J.~W.~F.~Valle,
  JHEP {\bf 0309}, 071 (2003)
  [arXiv:hep-ph/0304232].

\bibitem{Kaplan:2007ap}
  D.~E.~Kaplan and K.~Rehermann,
  JHEP {\bf 0710}, 056 (2007)
  [arXiv:0705.3426 [hep-ph]].

\bibitem{Dedes:2006ni}
  A.~Dedes, S.~Rimmer and J.~Rosiek,
  JHEP {\bf 0608}, 005 (2006)
  [arXiv:hep-ph/0603225].

\bibitem{Diaz:2004fu}
  M.~A.~D\'{\i}az, C.~Mora and A.~R.~Zerwekh,
  Eur.\ Phys.\ J.\  C {\bf 44}, 277 (2005)
  [arXiv:hep-ph/0410285].

\bibitem{Maltoni:2004ei}
M.~Maltoni, T.~Schwetz, M.~A. Tortola, and J.~W.~F. Valle,
\newblock New J. Phys. {\bf 6}, 122 (2004),
\newblock version 6 of the arXiv, hep-ph/0405172, provides results updated as
  of September 2007; previous works by other groups as well as the relevant
  experimental references are given therein.

\bibitem{Porod:2003um}
  W.~Porod,
  Comput.\ Phys.\ Commun.\  {\bf 153}, 275 (2003)
  [arXiv:hep-ph/0301101].


\bibitem{Skands:2003cj}
P.~Skands {\em et~al.},
\newblock JHEP {\bf 07}, 036 (2004), [hep-ph/0311123].


\bibitem{Sjostrand:2000wi}
T.~Sj\"ostrand {\em et~al.},
\newblock Comput. Phys. Commun. {\bf 135}, 238 (2001), hep-ph/0010017.

\bibitem{Sjostrand:1993yb}
T.~Sj\"ostrand,
\newblock Comput. Phys. Commun. {\bf 82}, 74 (1994).

%
\bibitem{Hirsch:2003fe}
  M.~Hirsch and W.~Porod,
  Phys.\ Rev.\  D {\bf 68}, 115007 (2003)
  [arXiv:hep-ph/0307364].
\end{thebibliography}

\end{document}